\documentclass[twocolumn,floatfix,showpacs,prd,aps,tightenlines]{revtex4}
\usepackage{graphicx}
\usepackage{psfrag}
\usepackage{dcolumn}
\usepackage{bm}

\begin{document}

\title{Spinning-black-hole binaries: The orbital hang up}

\author{M. Campanelli}  \affiliation{Department of Physics and Astronomy,
and Center for Gravitational Wave Astronomy,
The University of Texas at Brownsville, Brownsville, Texas 78520}

\author{C. O. Lousto} \affiliation{Department of Physics and Astronomy,
and Center for Gravitational Wave Astronomy,
The University of Texas at Brownsville, Brownsville, Texas 78520}

\author{Y. Zlochower} \affiliation{Department of Physics and Astronomy,
and Center for Gravitational Wave Astronomy,
The University of Texas at Brownsville, Brownsville, Texas 78520}

\date{\today}

\begin{abstract}
We present the first fully-nonlinear numerical study of the dynamics
of highly spinning black-hole binaries. We evolve binaries from
quasicircular orbits (as inferred from Post-Newtonian theory), and
find that the last stages of the orbital motion of black-hole binaries
are profoundly affected by their individual spins. In order to cleanly
display its effects, we consider two equal mass holes with individual
spin parameters $S/m^2=0.757$, both aligned and anti-aligned with the
orbital angular momentum (and compare with the spinless case), and
with an initial orbital period of $125M$. We find that the aligned
case completes three orbits and  merges significantly after the
anti-aligned case, which completes less than one orbit. The total
energy radiated for the former case is $\approx7\%$ while for the
latter it is only $\approx2\%$. The final Kerr hole remnants have
rotation parameters $a/M=0.89$ and $a/M=0.44$ respectively, showing
the unlikeliness of creating a maximally rotating black hole out of
the merger of two spinning holes.
\end{abstract}

 \maketitle

\section{Introduction}\label{Sec:Intro}

Spinning black holes play an important role in some of the most
energetic astrophysical phenomena in the universe. They form part of
the main engine of gamma-ray bursts, being much more efficient at
converting matter into radiation than non-spinning black holes. They
are also responsible for the radio jets observed in active galactic
nuclei, and the merger of two non-aligned spinning black holes is the
likely explanation for the rapid directional changes observed in these
jets when galaxies collide~\cite{Merritt:2002hc}.
Recent estimates~\cite{Shafee:2005ef} of the spin of stellar mass
black holes by spectral fitting of the X-Ray continuum set the
rotation parameter of two dynamically confirmed black holes at
$a/M\sim0.75$.  Accretion, of course, can spin up black holes,
reaching up to a sub-maximal spin rate of $a/M\sim0.95$, when
magneto-hydrodynamics is taken into account~\cite{Gammie:2003qi,
Shapiro:2004ud}. Other models using the combined effects of gas
accretion and binary-black-hole coalescence suggest that black
holes may be rapidly rotating in all epochs~\cite{Volonteri:2004cf}.

Recently new numerical
techniques to solve the field equations of General Relativity have
been developed~\cite{Bruegmann:2003aw,Pretorius:2005gq,Diener:2005mg,Campanelli:2005dd,Baker:2005vv}
that make it possible to stably evolve black-hole binaries for several
orbits and to compute the corresponding gravitational waveforms
~\cite{Campanelli:2006gf,Baker:2006yw,Pretorius:2006tp}. Numerical
simulations of unequal-mass black-hole binaries, along with the
calculation of the merger kicks, have been reported in
Refs~\cite{Campanelli:2004zw, Herrmann:2006ks, Baker:2006vn}.  While
research has been mainly focused on initially non-spinning black
holes, there are important questions to be addressed when we consider
highly-spinning black holes (see~\cite{Flanagan97a} and references
therein). 

In this paper we study how the emission of gravitational
radiation affects the orbital trajectory of highly-spinning,
equal-mass black holes as a function of the spin orientation.
In order to maximize
the effect, we consider black-hole binaries with both spins aligned
and anti-aligned with the orbital angular momentum, as well as the 
corresponding spinless case. We
shall consider quasicircular orbit initial data with the same initial
orbital period (as determined by the third post-Newtonian (3PN)
expansion). In this way differences in the subsequent evolution can be
attributed to the differences in the generation and emission of
gravitational radiation.

In Ref.~\cite{Baker:2004wv} the numerical evolutions of spinning binaries
were studied for relatively modest values of the spins
($-0.25\leq S/m^2\leq 0.17$, $m$ being the horizon mass of the
individual holes) using the `Lazarus' technique of matching full
numerical evolutions to perturbation theory. In those evolutions the
spin of the remnant Kerr hole increased with $S/M^2$ for
the aligned case. Extrapolation to maximally spinning individual holes
indicated that the remnant would remain sub-maximal for
$S/m^2<0.85$. We will revisit this scenario, now reaching much higher
values of the individual spins in order to make a more accurate
statement.

\section{Initial data}\label{Sec:Init}

\begin{table}
\caption{Initial data for quasicircular orbits of black-hole binaries
with spin. The holes have proper horizon separation $l$, with
puncture locations $(0,\pm y,0)$, linear momenta $(\mp P, 0, 0)$, and
spin $(0,0,S)$.  $J$ is the total angular momentum, $L$ is the orbital
angular momentum, $\Omega$ is the orbital frequency, $m_p$ is the
individual puncture mass.
All in
units of the ADM mass $M$.}
\begin{ruledtabular}
\begin{tabular}{llll}\label{table:ID}
$S/m^2$       &++0.757   &  0.00  &-\,-\,0.757  \\
\hline
$l/M$     & 9.27   &  9.62  & 10.34 \\
$y/M$     & 3.0595 & 3.280   & 3.465   \\
$P/M$     & 0.1291  & 0.1336  & 0.1382  \\
$S/M^2$   &+0.1939 & 0.000   &-0.1924 \\
$J/M^2$   & 1.1778    & 0.8764  & 0.5729   \\
$L/M^2$   & 0.7900   & 0.8764  & 0.9577   \\
$M\Omega$ & 0.0500  & 0.0500 & 0.0500  \\
$m_p/M$     & 0.3344    & 0.4851  & 0.3344   \\

\end{tabular}
\end{ruledtabular}
\end{table}

We use the Brandt-Br\"ugmann puncture approach along with the elliptic
solver BAM\_Elliptic~\cite{Brandt97b,cactus_web} to compute initial
data. Table~\ref{table:ID} gives our choice of initial parameters. We
have taken a fiducial angular frequency of $M\Omega=0.05$, which
corresponds to an orbital period of approximately $T=125M$. This, accordingly to
our previous simulation for non-spinning black
holes~\cite{Campanelli:2006gf} makes the binary complete more than a
full orbit before the black holes merge. We choose individual spins
$S=\pm0.757\,m^2$ (as measured using isolated horizon techniques~\cite{Dreyer02a})
to guarantee that the total angular momentum in the
aligned case exceeds $M^2$, the maximum allowed
value for a common horizon to form. The gravitational radiation
emitted should efficiently carry out angular momentum from the system
in order for the {\it cosmic censorship} conjecture to hold
\cite{Wald84}. We can thus begin to explore its validity here, and
this will be the subject of a more detailed study in a forthcoming
paper by the authors.

With our choices of the spins and the orbital angular frequency, we
determine the remaining orbital parameters by imposing quasicircular
orbits according to the second post-Newtonian expansion of spinning
particles~\cite{Kidder:1995zr} extended by the third-order orbital
corrections~\cite{Blanchet:1999pm}. We then give those parameters to
the exact Bowen~\cite{Bowen79} solution of the momentum constraints and solve for the
conformal factor of the (conformally flat) three-metric to complete
our choice of the initial data.
This post-Newtonian data should produce orbits with acceptably small
eccentricities, as can be seen when comparing the zero-spin
parameters in table~\ref{table:ID} with others proposed in the
literature (e.g.~\cite{Tichy:2003qi}).
 We have also briefly studied the effects of
a different choice of the form of the initial data (Kerr conformal
extrinsic curvature) for spinning black holes, as proposed in
Ref.~\cite{Dain:2002ee}. However, the spurious radiation in the
initial data is dominated by the momentum terms and both data sets
give
nearly identical waveforms (see Fig.~\ref{fig:bspp_wave}).

\section{Techniques}\label{Sec:techniques}

We evolved these black-hole-binary data sets using the {\it
LazEv}~\cite{Zlochower:2005bj} implementation of the moving puncture
approach~\cite{Campanelli:2005dd, Baker:2005vv}.  In our version of
the moving puncture approach~\cite{Campanelli:2005dd} we replace the
BSSN~\cite{Nakamura87,Shibata95, Baumgarte99} conformal exponent
$\phi$, which is infinite on the punctures, with the initially
$C^4$ field $\chi
= \exp(-4\phi)$.  This new variable, along with the other BSSN
variables, will remain finite provided that one uses a suitable choice
for the gauge.

We obtained accurate, convergent waveforms by evolving this system in
conjunction with a modified 1+log lapse, a modified Gamma-driver shift
condition~\cite{Alcubierre02a,Campanelli:2005dd}, and an initial lapse
$\alpha\sim\psi_{BL}^{-4}$.  The lapse and shift are evolved with
$(\partial_t - \beta^i \partial_i) \alpha = - 2 \alpha K$, 
$\partial_t \beta^a = B^a$, and $\partial_t B^a = 3/4 \partial_t \tilde \Gamma^a - \eta B^a$.
These gauge conditions require careful treatment of $\chi$
near the puncture in order for the system to remain stable
\cite{Campanelli:2005dd,Campanelli:2006gf}.
For our version of the moving puncture approach, we find that the
product $\alpha \tilde A^{ij} \partial_j \phi$ has to be initially $C^4$ on the
puncture. In the spinning case, $\tilde A^{ij}$ is $O(r^3)$ on the
puncture, thus requiring that $\alpha \propto r^3$ to
maintain differentiability. We therefore choose an initial lapse
$\alpha \sim \psi_{BL}^{-4}$ (which is $O(r^4)$ and $C^4$ on the
puncture). In particular, $\alpha(t=0) = 2/(1+\psi_{BL}^{4})$
reproduces the isotropic Schwarzschild lapse at large distances from
the hole.  The initial values of $\beta^i$ and $B^i$ were set to zero.

The minimum resolution required to accurately model the
dynamics of the merger scales with $m_p$. We would expect satisfactory results for
a minimum resolution of $h=M/30$ (based on the non-spinning case,
where satisfactory results were obtained with $h=M/21$, and the ratio
of the puncture masses in the spinning and non-spinning cases), $M$
being the total ADM mass. However, the additional power of $1/r$
introduced to $K_{ij}$ because of spin, necessitates even higher resolution
(we estimate $M/40 - M/50$) to get highly accurate waveforms.

We exploited the Pi-rotational symmetry about the $z$-axis as well as
the reflection symmetry about the $xy$ plane to reduce the size of the
computational domain by one quarter.

\section{Results}\label{Sec:results}

\begin{figure}
\begin{center}
\includegraphics[width=3.3in]{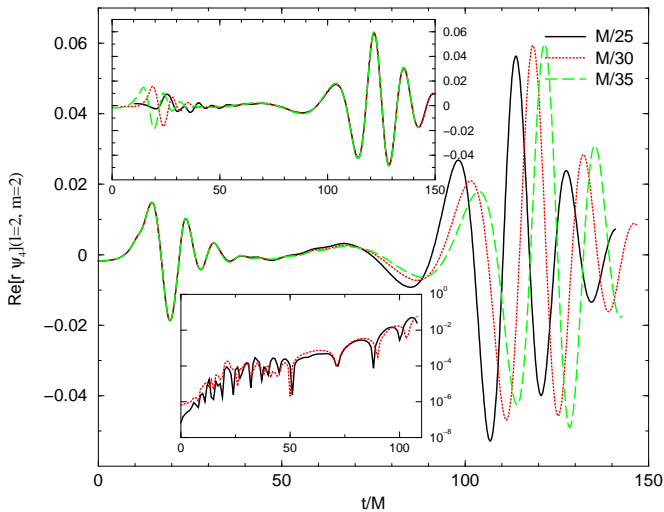}
\caption{The real $(\ell=2, m=2)$ component of $r\psi_4$ in the
Quasi-Kinnersley tetrad at $r=10M$ for the -\,-\,0.757 case. The lower inset
shows the differences $r\psi_4(M/25) - r\psi_4(M/30)$ (solid line) and
$r\psi_4(M/30) - r\psi_4(M/35)$ (dotted line), the latter rescaled by
$2.33$ to demonstrate fourth-order convergence. The lack of
convergence for $t<10M$ is due to roundoff effects in the initial data
solver.  The upper inset shows the real part of the phase-corrected
$(\ell=2,m=2)$ mode of $\psi_4$ at the same radius. Note the near-perfect
agreement after $t=45M$.}
\label{fig:bsmm_wave}
\end{center}
\end{figure}

We evolved the `-\,-\,0.757' configuration using grid sizes of
$320^2\times640$, $384^2\times768$, and $448^2\times896$ and
resolutions of $M/25$, $M/30$, and $M/35$ respectively. We used a
multiple transition fisheye~\cite{Campanelli:2006gf} to push the
physical boundaries to $134M$.  We calculate $\psi_4$ in the Quasi-Kinnersley
frame using the recently developed techniques of
Ref.~\cite{Campanelli:2005ia} that allow a meaningful extraction
closer to the hole.  In Fig.~\ref{fig:bsmm_wave} we show the real part
of the $(\ell=2,m=2)$ mode of $r\psi_4$ for the -\,-\,0.757 case
(extracted at $r=10M$) for these resolutions, as well as a convergence
plot of these data. The waveforms show fourth-order convergence up to
$t~\sim110M$. The phase error from the $h=M/25$ run becomes too large
to measure a meaningful convergence rate after $t\sim110M$.  Higher
resolution runs will remain convergent, as demonstrated by the better
phase agreement between the $M/30$ and $M/35$ runs.
 We extract the waveform at $10M$ to
minimize the effects of the extreme fisheye deresolution (which is
too strong in the far field to get accurate waveforms).  
After a time translation, the waveforms from the three runs only
differ by a constant phase factor. We calculate this factor and
plot the phase-corrected waveforms~\cite{Baker:2002qf,Baker:2006yw} 
in the upper inset of Fig.~\ref{fig:bsmm_wave}.
Note the near perfect agreement both in the orbital and plunge
waveforms. The waveforms calculated at $r=10M$ do not yield 
accurate estimates for the radiated energy (as expected, since
the observer is still in the near zone). However, as shown below,
we obtained highly accurate estimates for the radiation be examining
the remnant horizon. In order to obtain accurate measurements for the
 radiated energy and angular momentum from the waveform, one needs to
use a weaker fisheye deresolution in the outer region, and carefully
adjust the gauge so that the waveform is highly accurate at large
radii ($r\sim 50M$).

The relatively large phase errors in this spinning case compared with
our zero-spin simulations~\cite{Campanelli:2006gf} are due to the fact
that the effective resolution in the spinning case is smaller due to
the smaller value of $m_p$ as well as the lower order differentiability of the
spinning data compared to zero-spin. A likely explanation is that
numerical dissipation (which more strongly affects this higher
frequency data) causes the merger to happen sooner than expected. This
dissipation decreases with resolution.

We used Jonathan Thornburg's AHFinderDirect
thorn~\cite{Thornburg2003:AH-finding} to calculate the apparent
horizons. We find that the common horizon is first detected at
$t=105.5M$ and has a mass of $M_{\cal H} =0.978\pm.001M$ and rotation parameter of
$a/M_{\cal H} = 0.443\pm0.001$. 
During the merger $(2.2\pm0.1)\%$ of the mass and
$(26\pm2)\%$ of the angular momentum are converted into
radiation.

$M_{\cal H}$ and $a/M_{\cal H}$ were obtained
from the asymptotic values of the horizon surface area and the ratio of
the polar to equatorial circumferences (see 
Refs~\cite{Alcubierre:2004hr,Thornburg2003:AH-finding,Campanelli:2005dd}).
The ranges given
for these quantities arise from the uncertainties in obtaining
these values at finite time, and are independent of resolution.
Thus, these horizon parameters give an accurate and
 robust measurement (even
in the unresolved $h=M/25$ case) for the radiated mass and angular momentum.

\begin{figure}
\begin{center}
\includegraphics[width=2.7in]{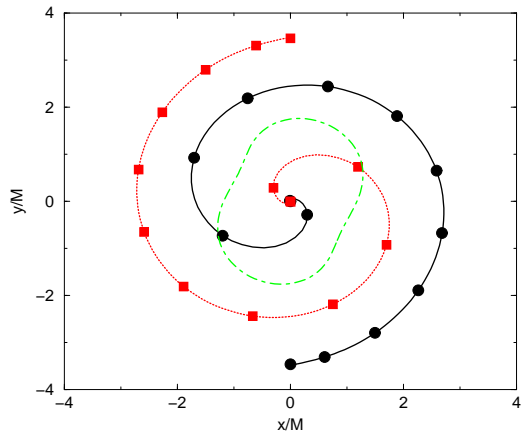}
\caption{The puncture trajectories on the $xy$ plane for the `-\,-' 
case with resolution $M/35$. The spirals are the puncture trajectories 
with ticks every 10M of evolution. The dot-dash `peanut shaped' figure is
the first detected common horizon at $105.5M$. The (extrapolated) 
period of the last orbit is around $120M$.  }
\label{fig:bsmm_track}
\end{center}
\end{figure}

Figure~\ref{fig:bsmm_track} shows the trajectories of the punctures
for the -\,-\,0.757 configuration as well as the projection of the
first common horizon on the $xy$ plane. It is evident from the
waveform and the track that the binary undergoes $\sim0.9$ orbits
before merging.

\begin{figure}
\begin{center}
\includegraphics[width=3.3in]{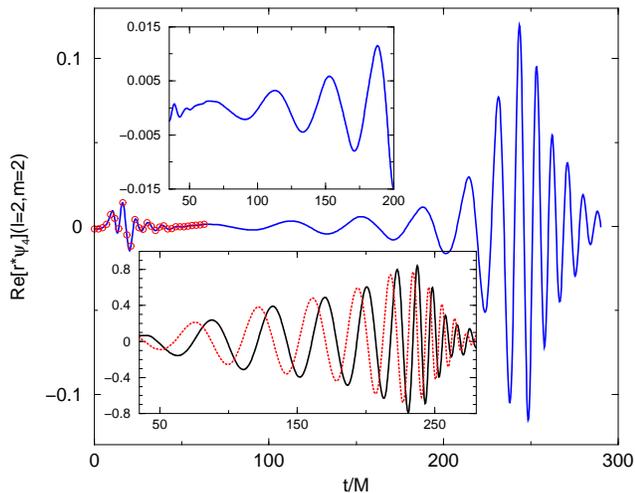}
\caption{The real part of the $(\ell=2,m=2)$ mode of $r\,\psi_4$ 
in the Quasi-Kinnersley frame at $r=10M$ from the `$++0.757$'
configuration. (The small circles are the early-time waveform from
conformal Kerr data.) The top inset shows a magnified view of the early orbital
motion. Note that the `++0.757' waveform has 6 wavelengths of orbital
motion prior to the plunge waveform (at $t\sim 232.5M$), indicating that
the binary orbited approximately three times before merging.
The bottom inset shows the real (solid) and imaginary (dotted) components of the
(2,2) component of the strain $h$ calculated at $r=10M$. }
\label{fig:bspp_wave}
\end{center}
\end{figure}

We evolved the `++' configuration with a grid size of $384^2\times768$
and resolution of $M/30$. We used multiple transition fisheye to push
the boundaries to $159M$.  In Fig.~\ref{fig:bspp_wave} we show the
$(\ell=2,m=2)$ mode of $r\psi_4$ in the Quasi-Kinnersley frame for the
`++0.757' case, again extracted at $r=10M$.
Note the `plunge' waveform is delayed by $\sim 120M$ compared to the
`-\,-\,0.757' case. The waveform shows approximately six periods of
orbital radiation prior to the plunge waveform, indicating that the
binary completed approximately three orbits.

We repeated the `++0.757' case with a gridsize of $448^2\times896$ and
resolution of $M/30$
to force the boundaries to $266M$. This new configuration allows us
to accurately obtain the horizon parameters (since they are
not contaminated by the boundary), but is too coarse in the far-field 
region to produce accurate waveforms.
The first common horizon was detected at
$t=232.5M$. In this case the final horizon had a mass of $0.933\pm.001M$
and spin of $0.890\pm.002$ (indicating that $(6.7\pm0.2\%)$ of the mass
and $(34\pm1)\%$ of the angular momentum are radiated away).
Table~\ref{table:results} gives a
summary of these results. Note that the above values for the radiated
energy are in rough agreement with those estimated using the effective 
one body approximation for maximally spinning holes~\cite{Buonanno:2005xu}.

\begin{figure}
\begin{center}
\includegraphics[width=2.7in]{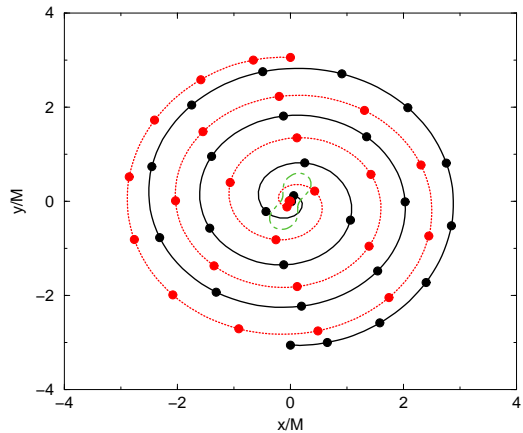}
\caption{The puncture trajectories on the $xy$ plane for `++' configuration
with resolution $M/30$. The spirals are the puncture trajectories with
ticks every 10M of evolution. The dot-dash `peanut shaped' figure is
the first detected common horizon at $t=232.5M$. The period of the
last orbit is around $36M$.  The last orbit begins when the punctures
are located at $1.4M$ from the origin (in these coordinates).  }
\label{fig:bspp_track}
\end{center}
\end{figure}

Figure~\ref{fig:bspp_track} shows the track for the `++0.757'
configuration.  Note that the spiral is much tighter than in the
`-\,-\,0.757' configuration, and that the binary completes roughly
$3$ orbits before the common horizon forms. Note also that the first
common horizon is much smaller in this case (in these coordinates)
than in the `-\,-\,0.757' case.

\begin{figure}
\begin{center}
\includegraphics[width=2.5in]{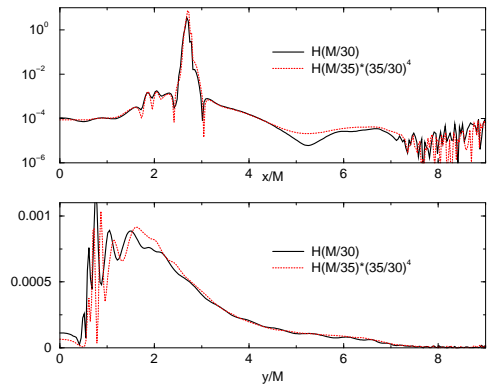}
\caption{The Hamiltonian constraint violation at $t=45M$ along the x-axis 
(top plot) and at $t=80M$ along the $y$-axis (bottom plot) 
for the $M/30$ and $M/35$ runs (the latter rescaled by $(35/30)^4$) for
the `-\,-' configuration. The punctures crossed the x-axis at $t=45M$ and
crossed the y-axis for the
second time at $t=75M$.  Note the reasonable fourth-order
convergence (except at the puncture). Points contaminated by 
boundary errors have been excluded
from the plot.  The high frequency violations near the numerical
coordinate $y/M=\pm9$ are due to the extreme fisheye deresolution near
the boundary, and converge with resolution.  }
\label{fig:HC_conv}
\end{center}
\end{figure}

To demonstrate consistency with the General Relativity field
equations, we calculated the Hamiltonian constrain violation. The
constraint converges to fourth-order outside a small region
surrounding the puncture (the constraint violation on the
nearest neighboring points to the puncture is roughly independent of
resolution, but this non-converging error does not propagate outside
the individual horizons).  Figure~\ref{fig:HC_conv} shows the
Hamiltonian constraint violations for the `-\,-' configuration along
the $x$-axes at $t=45M$ and along the $y$-axis at $t=80M$ 
(at the time when the punctures cross the $x$-axis and
$5M$ after the punctures cross
the $y$-axis for the second time) for the $M/30$ and $M/35$ runs.  The
constraint is convergent everywhere except points contaminated by
boundary errors (these points have been removed from the plot)
and at the points closest to the puncture.

We complete our initial study with the corresponding spinless case as 
a reference point. For details on the accuracy and evolution of the
spinless case see Ref.~\cite{Campanelli:2006gf}. 
We evolved the zero-spin case with a resolution of $M/22.5$ and gridsize
of $320^2\times640$ (the outer boundary was located at $216M$).
The first common horizon formed at $t=161M$ with mass $(0.965\pm.001)\%$ and
spin $a/M = (0.688\pm.001)$. This corresponds to a radiated
energy and angular momentum of $(3.5\pm0.1)\%$ and $(26.9\pm0.1)\%$
 respectively.

\begin{table}
\caption{Results of the evolution as determined from 
the remnant horizons. The horizon formed at $t=224.5M$ for the `++0.757'
configuration, $t=105.5M$ for the `-\,-\,0.757'
configuration, and $t=161M$ for the spinless configuration.}
\begin{ruledtabular}
\begin{tabular}{llll}\label{table:results}
$S/m^2$ & $E_{rad}/M_{ADM}$ & $J_{rad}/J_{ADM}$ & $a/M_{\cal H}$  \\
\hline
++0.757 & $(6.7\pm0.2)\%$ & $(34\pm1)\%$ & $0.890\pm0.002$ \\ 
\,-\,-\,0.757 & $(2.2\pm0.1)\%$ & $(26\pm2)\%$ &$0.443\pm0.001$ \\ 
\,\,\,\,\,\,0.00 & $(3.5\pm.1)\%$ & $(26.9\pm0.1)\%$ & $0.688\pm0.001$ \\
\end{tabular}
\end{ruledtabular}
\end{table}

\section{Discussion}\label{Sec:Dis}

In this paper we have shown that the `moving puncture' approach can be
used to accurately simulate the inspiral orbit of spinning-black-hole
binaries.  We found that the spin-orbit coupling delays the
onset of the plunge phase (compared to the non-spinning case) when the
spins are aligned with the orbital angular momentum, while
in the anti-aligned case the plunge phase is hastened. 
In all cases, the black holes  merge to form a single Kerr
black hole with rotation parameter $a/M < 1$.

A fit of the remnant spin to the spins of the initial
black holes leads to the simple extrapolation
formula $a/M\approx0.688+0.298(S/m^2)-0.038(S/m^2)^2$. Although more accurate
simulations are needed, these results show that it is very unlikely to
form a nearly maximally rotating black hole out of the merger of two
highly spinning ones.  Our results reinforce the same qualitative
conclusions reached with the Lazarus approach~\cite{Baker:2004wv},
and are consistent with those in Refs~\cite{Flanagan97a,Buonanno:2005xu}.
Extrapolation of the radiated energy 
to maximally rotating black holes
with the fit $E_{rad}/M\approx3.5+3(S/m^2)+152/90(S/m^2)^2$
set it near $8.2\%$, not far from the assumed $10\%$ in 
Ref.~\cite{Flanagan97a} a decade ago.

For the `-\,-\,0.757' case the final orbit lasts $\sim 120M$
starting at a separation of $7M$ in coordinate space, while for the
`++0.757' case we have found that the duration of the last orbit is
$\sim 36M$ at a coordinate separation of $2.8M$. It is worth noting
from~\cite{Pfeiffer:2000um}, that the orbital period associated with the ISCO for the `++0.17' case
is roughly $27M$ at a coordinate separation of $1.6M$, and that no
ISCO was found for higher spin configurations. This highlights again the importance of the
gravitational radiation in the late binary black hole dynamics which 
is not captured in the determinations of the ISCO.
On the other hand, the dependence of the ISCO on spin correctly implies
that the `++' configurations are more stable at close separations
than the `-\,-' configurations.  This stability property is observed in the
significantly tighter spiral displayed in Fig.~\ref{fig:bspp_track}.

The Post-Newtonian equations of motion~\cite{Kidder:1995zr} indicate
that the leading spin-orbit interaction is of 1.5PN order, while the
spin-spin interaction is of 2PN order. It is the spin-orbit
interaction (attractive/repulsive for -\,-/++ configurations respectively)
responsible for the longer the stability of the aligned
spin binary.

Many outstanding issues involving spinning black hole scenarios remain
to be explored. We plan to study some of them next, including
additional values of the individual spins for the `++' cases in order
to better extrapolate the results to the limiting maximally rotating
individual holes, as well as unaligned spins to study precessional effects.
Finally, more
significant computer resources and the use of Adaptive Mesh Refinement
techniques combined with higher order finite difference methods will
be needed to achieve the accuracy required to aid gravitational
wave detection efforts~\cite{Miller:2005qu}.

\acknowledgments
We thank Bernard Kelly for careful reading of this text.
We thank Erik Schnetter for providing the Cactus thorns to implement
Pi-symmetry boundary conditions. We thank the referee for helpful
suggestions. We gratefully acknowledge the
support of the NASA Center for Gravitational Wave Astronomy at
University of Texas at Brownsville (NAG5-13396) and the NSF for
financial support from grants PHY-0140326 and
PHY-0354867. Computational resources were performed by the 70-node
`Funes' cluster at UTB.
	
\bibliographystyle{apsrev}
\bibliography{../../Lazarus/bibtex/references}

\begin{thebibliography}{39}
\expandafter\ifx\csname natexlab\endcsname\relax\def\natexlab#1{#1}\fi
\expandafter\ifx\csname bibnamefont\endcsname\relax
  \def\bibnamefont#1{#1}\fi
\expandafter\ifx\csname bibfnamefont\endcsname\relax
  \def\bibfnamefont#1{#1}\fi
\expandafter\ifx\csname citenamefont\endcsname\relax
  \def\citenamefont#1{#1}\fi
\expandafter\ifx\csname url\endcsname\relax
  \def\url#1{\texttt{#1}}\fi
\expandafter\ifx\csname urlprefix\endcsname\relax\def\urlprefix{URL }\fi
\providecommand{\bibinfo}[2]{#2}
\providecommand{\eprint}[2][]{\url{#2}}

\bibitem[{\citenamefont{Merritt and Ekers}(2002)}]{Merritt:2002hc}
\bibinfo{author}{\bibfnamefont{D.}~\bibnamefont{Merritt}} \bibnamefont{and}
  \bibinfo{author}{\bibfnamefont{R.~D.} \bibnamefont{Ekers}},
  \bibinfo{journal}{Science} \textbf{\bibinfo{volume}{297}},
  \bibinfo{pages}{1310} (\bibinfo{year}{2002}).

\bibitem[{\citenamefont{Shafee et~al.}(2006)}]{Shafee:2005ef}
\bibinfo{author}{\bibfnamefont{R.}~\bibnamefont{Shafee}} \bibnamefont{et~al.},
  \bibinfo{journal}{Astrophys. J.} \textbf{\bibinfo{volume}{636}},
  \bibinfo{pages}{L113} (\bibinfo{year}{2006}).

\bibitem[{\citenamefont{Shapiro}(2005)}]{Shapiro:2004ud}
\bibinfo{author}{\bibfnamefont{S.~L.} \bibnamefont{Shapiro}},
  \bibinfo{journal}{Astrophys. J.} \textbf{\bibinfo{volume}{620}},
  \bibinfo{pages}{59} (\bibinfo{year}{2005}).

\bibitem[{\citenamefont{Gammie et~al.}(2004)\citenamefont{Gammie, Shapiro, and
  McKinney}}]{Gammie:2003qi}
\bibinfo{author}{\bibfnamefont{C.~F.} \bibnamefont{Gammie}},
  \bibinfo{author}{\bibfnamefont{S.~L.} \bibnamefont{Shapiro}},
  \bibnamefont{and} \bibinfo{author}{\bibfnamefont{J.~C.}
  \bibnamefont{McKinney}}, \bibinfo{journal}{Astrophys. J.}
  \textbf{\bibinfo{volume}{602}}, \bibinfo{pages}{312} (\bibinfo{year}{2004}).

\bibitem[{\citenamefont{Volonteri et~al.}(2005)\citenamefont{Volonteri, Madau,
  Quataert, and Rees}}]{Volonteri:2004cf}
\bibinfo{author}{\bibfnamefont{M.}~\bibnamefont{Volonteri}},
  \bibinfo{author}{\bibfnamefont{P.}~\bibnamefont{Madau}},
  \bibinfo{author}{\bibfnamefont{E.}~\bibnamefont{Quataert}}, \bibnamefont{and}
  \bibinfo{author}{\bibfnamefont{M.~J.} \bibnamefont{Rees}},
  \bibinfo{journal}{Astrophys. J.} \textbf{\bibinfo{volume}{620}},
  \bibinfo{pages}{69} (\bibinfo{year}{2005}).

\bibitem[{\citenamefont{Br\"ugmann et~al.}(2004)\citenamefont{Br\"ugmann,
  Tichy, and Jansen}}]{Bruegmann:2003aw}
\bibinfo{author}{\bibfnamefont{B.}~\bibnamefont{Br\"ugmann}},
  \bibinfo{author}{\bibfnamefont{W.}~\bibnamefont{Tichy}}, \bibnamefont{and}
  \bibinfo{author}{\bibfnamefont{N.}~\bibnamefont{Jansen}},
  \bibinfo{journal}{Phys. Rev. Lett.} \textbf{\bibinfo{volume}{92}},
  \bibinfo{pages}{211101} (\bibinfo{year}{2004}).

\bibitem[{\citenamefont{Diener et~al.}(2006)}]{Diener:2005mg}
\bibinfo{author}{\bibfnamefont{P.}~\bibnamefont{Diener}} \bibnamefont{et~al.},
  \bibinfo{journal}{Phys. Rev. Lett.} \textbf{\bibinfo{volume}{96}},
  \bibinfo{pages}{121101} (\bibinfo{year}{2006}).

\bibitem[{\citenamefont{Pretorius}(2005)}]{Pretorius:2005gq}
\bibinfo{author}{\bibfnamefont{F.}~\bibnamefont{Pretorius}},
  \bibinfo{journal}{Phys. Rev. Lett.} \textbf{\bibinfo{volume}{95}},
  \bibinfo{pages}{121101} (\bibinfo{year}{2005}).

\bibitem[{\citenamefont{Campanelli
  et~al.}(2006{\natexlab{a}})\citenamefont{Campanelli, Lousto, Marronetti, and
  Zlochower}}]{Campanelli:2005dd}
\bibinfo{author}{\bibfnamefont{M.}~\bibnamefont{Campanelli}},
  \bibinfo{author}{\bibfnamefont{C.~O.} \bibnamefont{Lousto}},
  \bibinfo{author}{\bibfnamefont{P.}~\bibnamefont{Marronetti}},
  \bibnamefont{and}
  \bibinfo{author}{\bibfnamefont{Y.}~\bibnamefont{Zlochower}},
  \bibinfo{journal}{Phys. Rev. Lett.} \textbf{\bibinfo{volume}{96}},
  \bibinfo{pages}{111101} (\bibinfo{year}{2006}{\natexlab{a}}).

\bibitem[{\citenamefont{Baker et~al.}(2006{\natexlab{a}})\citenamefont{Baker,
  Centrella, Choi, Koppitz, and van Meter}}]{Baker:2005vv}
\bibinfo{author}{\bibfnamefont{J.~G.} \bibnamefont{Baker}},
  \bibinfo{author}{\bibfnamefont{J.}~\bibnamefont{Centrella}},
  \bibinfo{author}{\bibfnamefont{D.-I.} \bibnamefont{Choi}},
  \bibinfo{author}{\bibfnamefont{M.}~\bibnamefont{Koppitz}}, \bibnamefont{and}
  \bibinfo{author}{\bibfnamefont{J.}~\bibnamefont{van Meter}},
  \bibinfo{journal}{Phys. Rev. Lett.} \textbf{\bibinfo{volume}{96}},
  \bibinfo{pages}{111102} (\bibinfo{year}{2006}{\natexlab{a}}).

\bibitem[{\citenamefont{Campanelli
  et~al.}(2006{\natexlab{b}})\citenamefont{Campanelli, Lousto, and
  Zlochower}}]{Campanelli:2006gf}
\bibinfo{author}{\bibfnamefont{M.}~\bibnamefont{Campanelli}},
  \bibinfo{author}{\bibfnamefont{C.~O.} \bibnamefont{Lousto}},
  \bibnamefont{and}
  \bibinfo{author}{\bibfnamefont{Y.}~\bibnamefont{Zlochower}},
  \bibinfo{journal}{Phys. Rev. D} \textbf{\bibinfo{volume}{73}},
  \bibinfo{pages}{061501(R)} (\bibinfo{year}{2006}{\natexlab{b}}).

\bibitem[{\citenamefont{Baker et~al.}(2006{\natexlab{b}})\citenamefont{Baker,
  Centrella, Choi, Koppitz, and van Meter}}]{Baker:2006yw}
\bibinfo{author}{\bibfnamefont{J.~G.} \bibnamefont{Baker}},
  \bibinfo{author}{\bibfnamefont{J.}~\bibnamefont{Centrella}},
  \bibinfo{author}{\bibfnamefont{D.-I.} \bibnamefont{Choi}},
  \bibinfo{author}{\bibfnamefont{M.}~\bibnamefont{Koppitz}}, \bibnamefont{and}
  \bibinfo{author}{\bibfnamefont{J.}~\bibnamefont{van Meter}},
  \bibinfo{journal}{Phys. Rev. D} \textbf{\bibinfo{volume}{73}},
  \bibinfo{pages}{104002} (\bibinfo{year}{2006}{\natexlab{b}}).

\bibitem[{\citenamefont{Pretorius}(2006)}]{Pretorius:2006tp}
\bibinfo{author}{\bibfnamefont{F.}~\bibnamefont{Pretorius}}
  (\bibinfo{year}{2006}), \eprint{gr-qc/0602115}.

\bibitem[{\citenamefont{Campanelli}(2005)}]{Campanelli:2004zw}
\bibinfo{author}{\bibfnamefont{M.}~\bibnamefont{Campanelli}},
  \bibinfo{journal}{Class. Quant. Grav.} \textbf{\bibinfo{volume}{22}},
  \bibinfo{pages}{S387} (\bibinfo{year}{2005}).

\bibitem[{\citenamefont{Herrmann et~al.}(2006)\citenamefont{Herrmann,
  Shoemaker, and Laguna}}]{Herrmann:2006ks}
\bibinfo{author}{\bibfnamefont{F.}~\bibnamefont{Herrmann}},
  \bibinfo{author}{\bibfnamefont{D.}~\bibnamefont{Shoemaker}},
  \bibnamefont{and} \bibinfo{author}{\bibfnamefont{P.}~\bibnamefont{Laguna}}
  (\bibinfo{year}{2006}), \eprint{gr-qc/0601026}.

\bibitem[{\citenamefont{Baker et~al.}(2006{\natexlab{c}})}]{Baker:2006vn}
\bibinfo{author}{\bibfnamefont{J.~G.} \bibnamefont{Baker}} \bibnamefont{et~al.}
  (\bibinfo{year}{2006}{\natexlab{c}}), \eprint{astro-ph/0603204}.

\bibitem[{\citenamefont{\'{E}anna \'{E}.~Flanagan and
  Hughes}(1998)}]{Flanagan97a}
\bibinfo{author}{\bibnamefont{\'{E}anna \'{E}.~Flanagan}} \bibnamefont{and}
  \bibinfo{author}{\bibfnamefont{S.~A.} \bibnamefont{Hughes}},
  \bibinfo{journal}{Phys. Rev. D} \textbf{\bibinfo{volume}{57}},
  \bibinfo{pages}{4535} (\bibinfo{year}{1998}).

\bibitem[{\citenamefont{Baker et~al.}(2004)\citenamefont{Baker, Campanelli,
  Lousto, and Takahashi}}]{Baker:2004wv}
\bibinfo{author}{\bibfnamefont{J.}~\bibnamefont{Baker}},
  \bibinfo{author}{\bibfnamefont{M.}~\bibnamefont{Campanelli}},
  \bibinfo{author}{\bibfnamefont{C.~O.} \bibnamefont{Lousto}},
  \bibnamefont{and}
  \bibinfo{author}{\bibfnamefont{R.}~\bibnamefont{Takahashi}},
  \bibinfo{journal}{Phys. Rev. D} \textbf{\bibinfo{volume}{69}},
  \bibinfo{pages}{027505} (\bibinfo{year}{2004}).

\bibitem[{\citenamefont{Brandt and Br{\"u}gmann}(1997)}]{Brandt97b}
\bibinfo{author}{\bibfnamefont{S.}~\bibnamefont{Brandt}} \bibnamefont{and}
  \bibinfo{author}{\bibfnamefont{B.}~\bibnamefont{Br{\"u}gmann}},
  \bibinfo{journal}{Phys. Rev. Lett.} \textbf{\bibinfo{volume}{78}},
  \bibinfo{pages}{3606} (\bibinfo{year}{1997}).

\bibitem[{cactus\_web()}]{cactus_web}
{\tt http://www.cactuscode.org}.
\bibitem[{\citenamefont{Dreyer et~al.}(2002)\citenamefont{Dreyer, Krishnan,
  Shoemaker, and Schnetter}}]{Dreyer02a}
\bibinfo{author}{\bibfnamefont{O.}~\bibnamefont{Dreyer}},
  \bibinfo{author}{\bibfnamefont{B.}~\bibnamefont{Krishnan}},
  \bibinfo{author}{\bibfnamefont{D.}~\bibnamefont{Shoemaker}},
  \bibnamefont{and}
  \bibinfo{author}{\bibfnamefont{E.}~\bibnamefont{Schnetter}},
  \bibinfo{journal}{Phys. Rev. D} \textbf{\bibinfo{volume}{67}},
  \bibinfo{pages}{024018} (\bibinfo{year}{2002}).

\bibitem[{\citenamefont{Wald}(1984)}]{Wald84}
\bibinfo{author}{\bibfnamefont{R.~M.} \bibnamefont{Wald}},
  \emph{\bibinfo{title}{General Relativity}} (\bibinfo{publisher}{The
  University of Chicago Press}, \bibinfo{address}{Chicago},
  \bibinfo{year}{1984}), ISBN \bibinfo{isbn}{0-226-87032-4 (hardcover),
  0-226-87033-2 (paperback)}.

\bibitem[{\citenamefont{Kidder}(1995)}]{Kidder:1995zr}
\bibinfo{author}{\bibfnamefont{L.~E.} \bibnamefont{Kidder}},
  \bibinfo{journal}{Phys. Rev. D} \textbf{\bibinfo{volume}{52}},
  \bibinfo{pages}{821} (\bibinfo{year}{1995}).

\bibitem[{\citenamefont{Blanchet}(1999)}]{Blanchet:1999pm}
\bibinfo{author}{\bibfnamefont{L.}~\bibnamefont{Blanchet}},
  \bibinfo{journal}{Pramana} \textbf{\bibinfo{volume}{53}}, \bibinfo{pages}{1}
  (\bibinfo{year}{1999}).

\bibitem[{\citenamefont{Bowen}(1979)}]{Bowen79}
\bibinfo{author}{\bibfnamefont{J.~M.} \bibnamefont{Bowen}},
  \bibinfo{journal}{Gen. Rel. Grav.} \textbf{\bibinfo{volume}{11}},
  \bibinfo{pages}{227} (\bibinfo{year}{1979}).

\bibitem[{\citenamefont{Tichy and Br{\"u}gmann}(2004)}]{Tichy:2003qi}
\bibinfo{author}{\bibfnamefont{W.}~\bibnamefont{Tichy}} \bibnamefont{and}
  \bibinfo{author}{\bibfnamefont{B.}~\bibnamefont{Br{\"u}gmann}},
  \bibinfo{journal}{Phys. Rev. D} \textbf{\bibinfo{volume}{69}},
  \bibinfo{pages}{024006} (\bibinfo{year}{2004}).

\bibitem[{\citenamefont{Dain et~al.}(2002)\citenamefont{Dain, Lousto, and
  Takahashi}}]{Dain:2002ee}
\bibinfo{author}{\bibfnamefont{S.}~\bibnamefont{Dain}},
  \bibinfo{author}{\bibfnamefont{C.~O.} \bibnamefont{Lousto}},
  \bibnamefont{and}
  \bibinfo{author}{\bibfnamefont{R.}~\bibnamefont{Takahashi}},
  \bibinfo{journal}{Phys. Rev. D} \textbf{\bibinfo{volume}{65}},
  \bibinfo{pages}{104038} (\bibinfo{year}{2002}).

\bibitem[{\citenamefont{Zlochower et~al.}(2005)\citenamefont{Zlochower, Baker,
  Campanelli, and Lousto}}]{Zlochower:2005bj}
\bibinfo{author}{\bibfnamefont{Y.}~\bibnamefont{Zlochower}},
  \bibinfo{author}{\bibfnamefont{J.~G.} \bibnamefont{Baker}},
  \bibinfo{author}{\bibfnamefont{M.}~\bibnamefont{Campanelli}},
  \bibnamefont{and} \bibinfo{author}{\bibfnamefont{C.~O.}
  \bibnamefont{Lousto}}, \bibinfo{journal}{Phys. Rev. D}
  \textbf{\bibinfo{volume}{72}}, \bibinfo{pages}{024021}
  (\bibinfo{year}{2005}).

\bibitem[{\citenamefont{Nakamura et~al.}(1987)\citenamefont{Nakamura, Oohara,
  and Kojima}}]{Nakamura87}
\bibinfo{author}{\bibfnamefont{T.}~\bibnamefont{Nakamura}},
  \bibinfo{author}{\bibfnamefont{K.}~\bibnamefont{Oohara}}, \bibnamefont{and}
  \bibinfo{author}{\bibfnamefont{Y.}~\bibnamefont{Kojima}},
  \bibinfo{journal}{Prog. Theor. Phys. Suppl.} \textbf{\bibinfo{volume}{90}},
  \bibinfo{pages}{1} (\bibinfo{year}{1987}).

\bibitem[{\citenamefont{Shibata and Nakamura}(1995)}]{Shibata95}
\bibinfo{author}{\bibfnamefont{M.}~\bibnamefont{Shibata}} \bibnamefont{and}
  \bibinfo{author}{\bibfnamefont{T.}~\bibnamefont{Nakamura}},
  \bibinfo{journal}{Phys. Rev. D} \textbf{\bibinfo{volume}{52}},
  \bibinfo{pages}{5428} (\bibinfo{year}{1995}).

\bibitem[{\citenamefont{Baumgarte and Shapiro}(1999)}]{Baumgarte99}
\bibinfo{author}{\bibfnamefont{T.~W.} \bibnamefont{Baumgarte}}
  \bibnamefont{and} \bibinfo{author}{\bibfnamefont{S.~L.}
  \bibnamefont{Shapiro}}, \bibinfo{journal}{Phys. Rev. D}
  \textbf{\bibinfo{volume}{59}}, \bibinfo{pages}{024007}
  (\bibinfo{year}{1999}).

\bibitem[{\citenamefont{Alcubierre et~al.}(2003)\citenamefont{Alcubierre,
  Br\"ugmann, Diener, Koppitz, Pollney, Seidel, and Takahashi}}]{Alcubierre02a}
\bibinfo{author}{\bibfnamefont{M.}~\bibnamefont{Alcubierre}},
  \bibinfo{author}{\bibfnamefont{B.}~\bibnamefont{Br\"ugmann}},
  \bibinfo{author}{\bibfnamefont{P.}~\bibnamefont{Diener}},
  \bibinfo{author}{\bibfnamefont{M.}~\bibnamefont{Koppitz}},
  \bibinfo{author}{\bibfnamefont{D.}~\bibnamefont{Pollney}},
  \bibinfo{author}{\bibfnamefont{E.}~\bibnamefont{Seidel}}, \bibnamefont{and}
  \bibinfo{author}{\bibfnamefont{R.}~\bibnamefont{Takahashi}},
  \bibinfo{journal}{Phys. Rev. D} \textbf{\bibinfo{volume}{67}},
  \bibinfo{pages}{084023} (\bibinfo{year}{2003}).

\bibitem[{\citenamefont{Campanelli
  et~al.}(2006{\natexlab{c}})\citenamefont{Campanelli, Kelly, and
  Lousto}}]{Campanelli:2005ia}
\bibinfo{author}{\bibfnamefont{M.}~\bibnamefont{Campanelli}},
  \bibinfo{author}{\bibfnamefont{B.}~\bibnamefont{Kelly}}, \bibnamefont{and}
  \bibinfo{author}{\bibfnamefont{C.~O.} \bibnamefont{Lousto}},
  \bibinfo{journal}{Phys. Rev. D} \textbf{\bibinfo{volume}{73}},
  \bibinfo{pages}{064005} (\bibinfo{year}{2006}{\natexlab{c}}).

\bibitem[{\citenamefont{Baker et~al.}(2002)\citenamefont{Baker, Campanelli,
  Lousto, and Takahashi}}]{Baker:2002qf}
\bibinfo{author}{\bibfnamefont{J.}~\bibnamefont{Baker}},
  \bibinfo{author}{\bibfnamefont{M.}~\bibnamefont{Campanelli}},
  \bibinfo{author}{\bibfnamefont{C.~O.} \bibnamefont{Lousto}},
  \bibnamefont{and}
  \bibinfo{author}{\bibfnamefont{R.}~\bibnamefont{Takahashi}},
  \bibinfo{journal}{Phys. Rev. D} \textbf{\bibinfo{volume}{65}},
  \bibinfo{pages}{124012} (\bibinfo{year}{2002}).

\bibitem[{\citenamefont{Thornburg}(2004)}]{Thornburg2003:AH-finding}
\bibinfo{author}{\bibfnamefont{J.}~\bibnamefont{Thornburg}},
  \bibinfo{journal}{Class. Quantum Grav.} \textbf{\bibinfo{volume}{21}},
  \bibinfo{pages}{743} (\bibinfo{year}{2004}).

\bibitem[{\citenamefont{Alcubierre et~al.}(2005)}]{Alcubierre:2004hr}
\bibinfo{author}{\bibfnamefont{M.}~\bibnamefont{Alcubierre}}
  \bibnamefont{et~al.}, \bibinfo{journal}{Phys. Rev. D}
  \textbf{\bibinfo{volume}{72}}, \bibinfo{pages}{044004}
  (\bibinfo{year}{2005}).

\bibitem[{\citenamefont{Buonanno et~al.}(2005)\citenamefont{Buonanno, Chen, and
  Damour}}]{Buonanno:2005xu}
\bibinfo{author}{\bibfnamefont{A.}~\bibnamefont{Buonanno}},
  \bibinfo{author}{\bibfnamefont{Y.}~\bibnamefont{Chen}}, \bibnamefont{and}
  \bibinfo{author}{\bibfnamefont{T.}~\bibnamefont{Damour}}
  (\bibinfo{year}{2005}), \eprint{gr-qc/0508067}.

\bibitem[{\citenamefont{Pfeiffer et~al.}(2000)\citenamefont{Pfeiffer,
  {T}eukolsky, and Cook}}]{Pfeiffer:2000um}
\bibinfo{author}{\bibfnamefont{H.~P.} \bibnamefont{Pfeiffer}},
  \bibinfo{author}{\bibfnamefont{S.~A.} \bibnamefont{{T}eukolsky}},
  \bibnamefont{and} \bibinfo{author}{\bibfnamefont{G.~B.} \bibnamefont{Cook}},
  \bibinfo{journal}{Phys. Rev. D} \textbf{\bibinfo{volume}{62}},
  \bibinfo{pages}{104018} (\bibinfo{year}{2000}).

\bibitem[{\citenamefont{Miller}(2005)}]{Miller:2005qu}
\bibinfo{author}{\bibfnamefont{M.}~\bibnamefont{Miller}},
  \bibinfo{journal}{Phys. Rev. D} \textbf{\bibinfo{volume}{71}},
  \bibinfo{pages}{104016} (\bibinfo{year}{2005}).

\end{thebibliography}
\thebibliography{rp}

\end{document}